\documentclass[amsmath,amssymb]{revtex4}

\usepackage[latin1]{inputenc}
\usepackage{graphicx}
\usepackage{dcolumn}
\usepackage{bm}
\usepackage{color}

\newcommand{\ket}[1]{|#1\rangle}

\begin{document}

\title{Quantum Multiplexing for Quantum Computer Networks.}
\author{Juan Carlos Garc\'ia-Escart\'in}
\email{juagar@tel.uva.es}
\author{Pedro Chamorro-Posada}
\affiliation{Departamento de Teor\'ia de la Se\~{n}al y Comunicaciones e Ingenier\'ia Telem\'atica. Universidad de Valladolid. Spain.}
\date{\today}
\begin{abstract}
In communication networks many different channels must share a limited amount of resources. In order to allow for multiple simultaneous communications, multiple access techniques are routinely employed. With quantum communication, it is possible to share a new kind of resource. All of the system channels can be accommodated into a single channel in a larger Hilbert space. In the scheme, a single line combines the information of all the users, and, at the receiver, the original quantum channels are recovered. The given multiplexer/demultiplexer circuit can perform this $n$ qubits to qudit transformation. Connections with superdense coding and classical multiple access schemes are discussed. 
\end{abstract}

\maketitle

\section{Introduction.}
Quantum information opens new perspectives in the fields of communication and computation \cite{NC00}. More efficient algorithms \cite{Sho97,Gro97} and secure cryptography \cite{BB84,Eke91} are only some of the applications of quantum mechanics to information technologies. 

As the physical realization of quantum computers comes closer, the interest in quantum computer networks is steadily growing. Small operational quantum key distribution networks have already been built \cite{Ell02,EPT03,Ell04,ECP05} and there are proposals for qubit teleportation networks compatible with the existing optical fibre infraestructure \cite{Sha02,YS03,LSF04}. Network elements are being generalized into the quantum case \cite{TK02,CW06}, and new applications like delayed commutation \cite{GC06b}, which uses quantum mechanics to perform tasks impossible with classical networks, have appeared. There are also advances in the related topic of quantum computer architecture, which treats the communication of the different inner blocks of a quantum computer \cite{OCC02,COI03}.

In multiuser networks, many users might want to share the channel, and multiple access issues arise. Quantum communication results that expand classical information transmission capacity, like superdense coding \cite{BW92} can also be applied to multiple access scenarios using higher-dimensional systems \cite{LLT02,GW02,FZL06}. There exist information theoretical results with bounds on the capacity of quantum multiple access channels transmitting classical information \cite{Win01}, including derivations for quantum optical channels \cite{CP04}. The classical information capacity of a quantum multiple access channel can be increased by quantum channel coding. There are practical optical schemes that exploit the superadditivity property of quantum coding, i.e. the ability to obtain an increase in the capacity  more than proportional to the size of the code, to improve the information transmission performance \cite{FTM03,BVF00}. With adequate quantum code design and decoding schemes, an increase in the word size, which can also be seen as the use of symbols from a higher-dimensional Hilbert space, allows to optimise the channel capacity, and even to adapt the channel capacity distribution between two senders \cite{HZH00}. Under this formulation, superdense coding can be defined and the properties of multiple-access bosonic channels have been studied \cite{YS04,YS05}.
 
In communication networks there is a limited amount of resources that the users have to share. It is in this framework where \emph{multiple access} techniques appear \cite{Skl83,Skl00}. Most communication systems employ one form or another of \emph{multiplexing}, i.e. the sharing of the channel by various users. In multiplexing, the information of many users is transmitted by a single channel. The information is transformed in order to reduce the use of the most scarce resource of the communication system of interest, and to take full advantage of the channel information capacity.

To avoid the interference between users, the signals in each subchannel must present some orthogonality properties in at least one domain. Common multiple access methods include time division multiple access, TDMA, where each user transmits in a different time slot so that their information does not overlap, frequency division multiple access, FDMA, where the signals are moved into different bands of the available spectrum, and code division multiple access, or CDMA where the algebraic properties of the signals allow to send them at the same time and in the same frequency range and to separate them at the receiver. Wavelength division multiple access, WDMA, the preferred method in optical networks, is a form of FDMA where the stress is put on the wavelength instead of on the related frequency parameter. 

Other multiple access methods are space division multiple access, SDMA, used mostly in wireless networks with adaptative antennas, and, polarization division multiple access, PDMA, where the separation of signals is made using orthogonal polarizations. 

These techniques can also be applied to quantum communication. A probabilistic SDMA scheme can be employed in passive optical networks \cite{Tow97,FGC05}. WDMA has been demonstrated in quantum key distribution networks \cite{BBG03,BBG04}, and in classical-quantum multiplexing, combining quantum key distribution with the transmission of classical data \cite{Tow97b,NTR05}. There are also non-optical multiplexing schemes, like the proposal for a magnetic version of FDMA in quantum communication with spin chains \cite{WLK06}. 

In this context, it is natural to consider the new resources quantum systems offer as possible candidates for new multiplexing schemes. The structure of the Hilbert space presents some interesting properties of orthogonality that can be exploited. In this paper, we will propose a scheme for Hilbert space division multiple access, HDMA, in quantum networks used for \emph{quantum information transmission}. In HDMA the information of many qubits is carried by a single element in a higher-dimensional Hilbert space. This d-dimensional information unit, or \emph{qudit}, will carry the same information as the original group of qubits. 

Section \ref{notation} introduces the gates and symbols used in the rest of the paper. Section \ref{QMUXGate} presents a Quantum Multiplexer Gate, and derives a general multiplexing scheme based on it. Section \ref{qubitqudit} compares the properties of bidimensional information units to those of d-dimensional systems. In section \ref{QST} a general quantum state transfer circuit for qudits is put forward. Section \ref{QMUX} derives a general multiplexer/demultiplexer scheme based on quantum state transfer. An example for a three channel multiplexer is given in section \ref{eg}. Finally, section \ref{discussion} discusses the results and examines quantum applications like superdense coding and classical multiplexing schemes in the light of the more general HDMA framework.

\section{Notation and gates.}
\label{notation}
We will work with qubits and qudits. Qubits are binary quantum information units that can exist in a superposition of states of the form $\ket{\psi}=\alpha \ket{0}+\beta \ket{1}$, where $|\alpha|^2$ and $|\beta|^2$ are the probabilities of finding $\ket{0}$ and $\ket{1}$ respectively. Qudits are d-dimensional quantum information units that, in the most general form, can be written as \[\ket{\psi}^d=\mathop{\sum}_{i=0}^{d-1}\alpha_i \ket{i}^d \textrm{, with} \sum_{i=0}^{d-1} \mid \!\alpha_i\! \mid^2 = 1.\] Qubits are the particular case for $d=2$. The coefficients $\alpha_i$ are complex numbers that give the probability of finding each value and also carry important phase information. The state of the qubits will be expressed in the usual ket notation whereas qudits will have an extra superscript to indicate the dimension of the Hilbert space, e.g. $\ket{\psi}$ for qubits, but $\ket{\psi}^d$ for a qudit state.

All the operations in the multiplexer will be given in terms of controlled NOT operations. For qubits the controlled NOT, CX, or CNOT, is an operator acting on a pair of qubits that preserves the state of the first qubit, the control qubit, and has at the output of the second qubit, or target, the logical XOR of both input qubits. It can also be seen as a modulo 2 addition: $CNOT\ket{x}\ket{y}=\ket{x}\ket{x\oplus y}=\ket{x}\ket{x + y \textrm{ (mod 2)}}$. This gate can be generalized for qudits as $CX^d=\ket{x}^d\ket{y}^d=\ket{x}^d\ket{x + y \textrm{ (mod d)}}^d$. Through the paper $\oplus$ will be used to indicate modulo $d$ addition, where $d$ can be determined from the superscript of the ket, and is taken to be 2 if nothing is specified. These operators can also be written in vectorial notation as $d^2 \times d^2$ unitary matrices.

The inverse operator of any quantum gate $U$ is its inverse matrix. For unitary evolution this inverse is $U^{\dag}$. For $CX^d$, ${CX^d}^{\dag}\ket{x}^d\ket{y}^d=\ket{x}^d\ket{x\ominus y}^d$, where $\ominus$ represents modulo $d$ subtraction. All the given qudit definitions hold for qubits ($d=2$), for which the CNOT operation is its own inverse.

The NOT gate, $X\ket{x}=\ket{x \oplus 1}$, can be generalized for qudits as $X_{m}^{d}\ket{x}^d=\ket{x\oplus m}^d$, for $m<d$. Thus, $X^2_{1}=NOT$. We will also define a new qudit control $C^{\mathcal{S}}U$, meaning that the operation $U$ will be applied on the target qudit if and only if the control qudit is in a state contained in the set $\mathcal{S}$. If no set is given, $\mathcal{S}=\{\ket{1}\}$ is supposed for two-dimensional systems, and the previous $\ket{x\oplus y}^d$ is kept in qudits. This way, CU recovers, for qubits, the meaning of a gate that is only applied when the control qubit is in $\ket{1}$. With this notation CNOT can also put as $CX^{2}_{1}$.

\section{Quantum Multiplexer Gate.}
\label{QMUXGate}
\begin{figure}[ht!]
\centering
\includegraphics{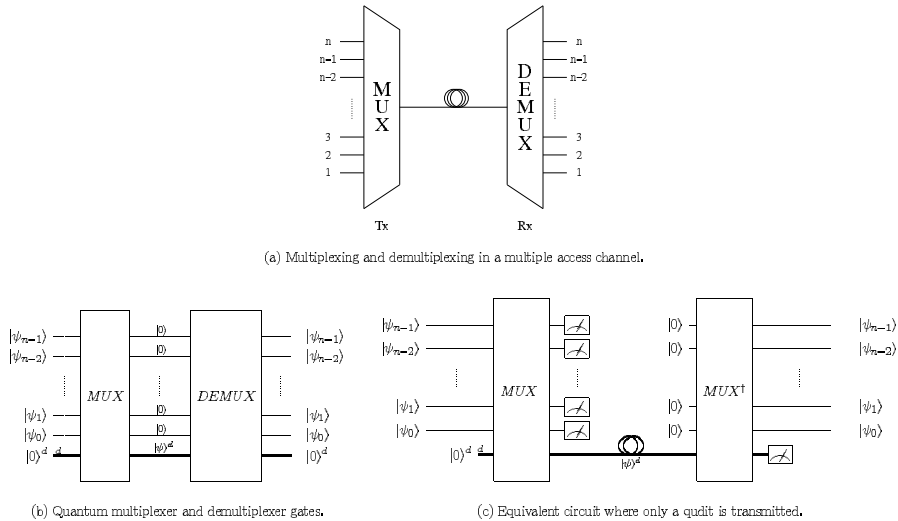}
\caption{Multiplexing schemes for classical and quantum communications.\label{muxgeneral}} 
\end{figure}

Multiplexers play an essential role in classical multiple access systems, as well as in many digital circuits. Multiplexers are blocks that take a group of input channels into a smaller number of outputs. The term multiplexer, or MUX, usually refers to the system that combines $n$ inputs into a single output. Combinations of this simpler multiplexer can give conversions between any desired number of inputs and outputs. The inverse operation, the expansion of one channel into the original $n$ signals, is performed by a demultiplexer, or DEMUX. Multiplexers and demultiplexers are usually represented as isosceles trapezoids with the longer side facing the $n$ lines, and the shorter side facing the single line. Figure \ref{muxgeneral}.a shows the usual configuration in classical communication systems. 

Similarly, a quantum MUX gate can be defined. There are a few restrictions, though. Quantum gates must be reversible. In the same way that quantum evolution alone does not allow to build a direct equivalent of the AND gate, an additional tool is needed to reduce the number of lines. Figure \ref{muxgeneral}.b gives a quantum multiplexer gate, QMUX, that transfers the state to a qudit and has at the qubit channels a fixed state, in this case $\ket{0}$. Then we apply a demultiplexer gate, $QDEMUX=QMUX^{\dag}$, that inverts the process. As the intermediate state of the $n$ qubits is known and there is no entanglement with the qudit, we can measure them on the transmitter side. Latter, at the receiver, we only need to generate $\ket{0}$ qubits to undo the transformation (Fig. \ref{muxgeneral}.c). This will be the architecture in our quantum multiplexer proposal.

\section{Qubits and qudits.}
\label{qubitqudit}
Qubits are the preferred quantum information unit. They are binary digital units encoded into two-dimensional systems. They are the generalization of the classical binary information unit, the bit. It is also possible to do computation with analogical variables \cite{How05}, or with multivalued logic \cite{Smi81,Hur84,Smi88}, with more than only two possible states. The first digital computers, like the ENIAC, were, in fact, decimal \cite{GG96}. There are also proposals for quantum computation with continuous variables \cite{SB99,BP00a,BV05}, and  qudits \cite{BGS02,BOB05,OBB06}, but, as occurs in the classical case, digital binary logic is the more widespread option. The simplicity of the gates and the binary logic intuition inherited from classical computers made it the starting point for the theoretical developments, and most of the physical implementation attempts have been made for qubits. There is no preferred multivalued logic to compete with qubits. Continuous variables proposals are better developed than qudit systems, but, as happens in classical computers, it is not likely that we will see in the next few years a multileveled or continuous system with the same degree of technological development and understanding we have of two-level systems.

Qudits can provide advantages from the point of view of communications. Their higher dimensionality gives a more compact mean of transmitting the same amount of information, increasing the information transmission rate. Furthermore, quantum communication protocols with qudits, like quantum coin tossing \cite{MVU05}, improve the performance of certain applications with respect to their qubit counterparts. Quantum cryptography using higher-dimensional alphabets \cite{BKB01} makes it easier to detect eavesdroppers \cite{BT00,WLA06}, is more robust agaisnt coherent attacks \cite{BM02,CBK02}, and, in general, brings a higher degree of security \cite{BP00b, BKB02}. Multilevel encoding also permits the incorporation of superdense coding schemes into secure direct communication \cite{WDL05}.  

It would be desirable to have at our disposal an element that can convert the qubits used for the local computations, with the better known qubit circuits, into flying qudits that take advantage of their superior communication properties. The quantum multiplexer will be defined in this context. The information of $n$ qubits will be carried in a $2^n$-dimensional qudit. A $2^n$-dimensional Hilbert space is the smallest that can describe the global state of $n$ arbitrary qubits. Imagine we use the qubits for classical information transmission. To be able to differenciate between the $2^n$ possible values $n$ bits have, we would need, at least, $2^n$ states. A system with a smaller number of states wouldn't allow for a deterministic recovery of the data, although a probabilistic coding can be obtained \cite{GWC06}.

\section{Quantum state transfer.}
\label{QST}
\begin{figure}[ht!]
\centering
\includegraphics{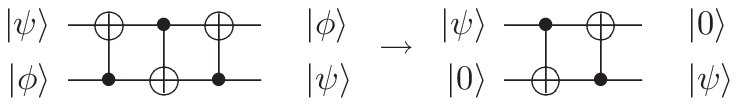}
\caption{Swap circuit.\label{SWAP}} 
\end{figure}

The transfer of information from one qubit to another is the starting point of many quantum information applications and is a primitive that appears in teleportation and superdense coding \cite{Mer01,Mer02,Mer03}. Particularly important is the quantum swap circuit (Fig. \ref{SWAP}, left), the quantum generalization of the classical reversible XOR swapping. In the figure, we see the usual representation for the CNOT gate, with a dot on the control qubit and a $\oplus$ on the target. The circuit acts in three stages $\ket{x}\ket{y}\rightarrow \ket{x\oplus y}\ket{x} \rightarrow \ket{x\oplus y}\ket{y\oplus x\oplus y} \rightarrow \ket{y}\ket{x}$. The same applies to superpositions. 

Figure \ref{SWAP} (right) shows how the circuit can be simplified if we are allowed to choose one of the input states, in this case choosing $\ket{0}$ for the second qubit. When controlled by $\ket{0}$, the CNOT operation is equivalent to the identity gate. The circuit on the right implements a quantum state transfer. The second CNOT gate is essential for a complete transfer. The first step already gives $\ket{x}\ket{x}$, but, without the last CNOT, the upper qubit is still entangled to the lower line and we have an entangling gate instead of a transfer in the proper sense. 

\begin{figure}[ht!]
\centering
\includegraphics{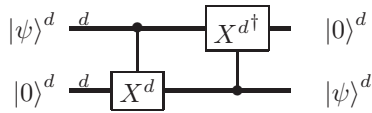}
\caption{Qudit state transfer circuit.\label{QuditST}} 
\end{figure}

This qubit state transfer can be extended to qudits generalizing the quantum swap circuit and performing a similar simplification (Fig. \ref{QuditST}). Here, qudits are represented as thicker lines with their dimension $d$ written at the beginning of the line. The control qudit is represented by a dot with a vertical line going to the controlled gate on the target qudit. 
 
\begin{figure}[ht!]
\centering
\includegraphics{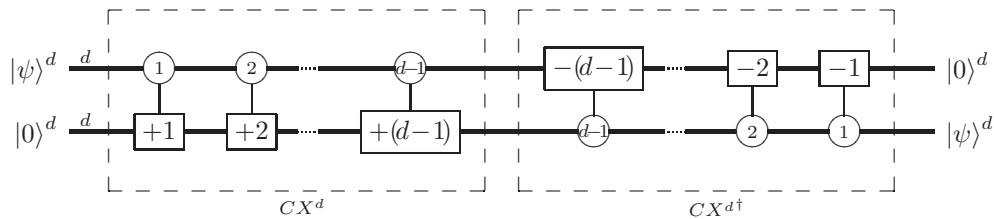}
\caption{Expanded qudit state transfer circuit. The +0 and -0 gates have no effect and can be omitted.\label{expQuditST}} 
\end{figure}

The $CX^d$ gate can be decomposed in a number of ways. The easiest to see is the concatenation of $C^{\mathcal{S}}X^d_x$ gates, each controlled by the corresponding $\ket{x}$ state, so that $CX^d\ket{x}\ket{y}=(\prod_{i=0}^{d-1}C^{\{\ket{i}\}}X^d_i)\ket{x}\ket{y}$. The control depends on different qudit states. So, each gate acts on a different subspace of the tensor product that gives the composite system's state $\ket{x}^d\ket{0}$ and the order of the gates is not important, i.e. they commute. In figure \ref{expQuditST} this qudit state control is represented by a circle containing the state number, and the $X^d_m$ gate by a $+m$ and the ${X^d_m}^{\dag}$ gate by a $-m$. Although, strictily speaking, only modulo d addition and subtraction give a unitary evolution, for HDMA simple addition and subtraction, without the circularity property, will suffice. This can be an advantage. In many physical systems qudits are implemented restricting to only a few states of higher-dimensional systems (even infinite-dimensional spaces) and addition and subtraction are easy while modular arithmetic is not. For instance, if the qudits are encoded in the orbital angular momentum of photons, the state number can be easily increased or decreased by a fixed number using holograms \cite{LPB02}, while there is no evident method for modulo $d$ operations. 

\begin{figure}[ht!]
\centering
\includegraphics{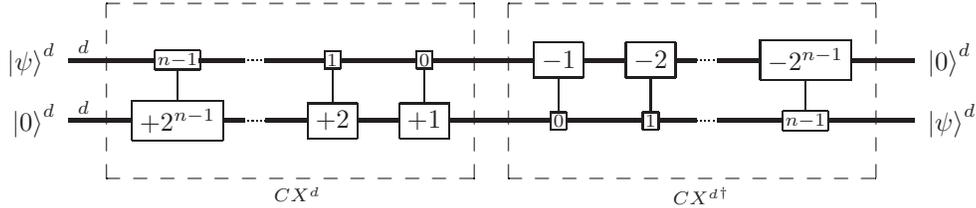}
\caption{Qudit state transfer with a binary expansion in the control.\label{binQuditST}} 
\end{figure}

In this application the main concern is on an \emph{n qubits to qudit state transfer}, and we will work with qudits such that $d=2^n$. In fact, $n$ qubits are a possible embodiment of a power of two-dimensional qudit. The multiplexer will convert this qudit into a more compact form. From that point of view, it is better to think in terms of a binary decomposition. For qudit state $\ket{x}$, with $x_{n-1}x_{n-2}\ldots x_{1}x_{0}$ the binary string for $x$, we will define the sets $\mathcal{S}_i=\{ \ket{x} : x $ div $ 2^i$ is odd $\}$, where a div b denotes the quotient of the division of a by b. The $C^{\mathcal{S}_i}X_{2^i}^d$ gate is a $X^d_{2^i}$ gate, controlled by the value of the $i$-th digit of the binary expression of the qudit state. Here, a state $\ket{x}^d$ acts on a distributed way, and the $+x$ sum is done by powers of two. After the $n$ steps the sums only happened when there was a 1 in the binary decomposition, and each $\ket{x}$ has done a $+x$ sum in the form $+x_{n-1}2^{n-1}+x_{n-2}2^{n-2}+\ldots+x_1 2 + x_0$. Fig. \ref{binQuditST} shows the quantum circuit using a square with the bit position index in the control qudit. 

This circuit gives a decomposition where the number of gates has been reduced from $2^n$ to $n$ and has an intuitive interpretation in terms of qubit controlled gates. The procedure can be translated to any base $l$ expansion and produce circuits for state transfer between $log_l d$ $l$-leveled systems and a qudit, $log_l d$ steps, for $d$ a power of $l$.  

\section{Quantum multiplexer.}
\label{QMUX}
At this point, all the elements of the quantum multiplexer are given. The scheme will consist of a $n$ qubits to qudit state transfer at the transmitter side, and the inverse qudit to $n$ qubits state transfer at the receiver side. The transfer will be done qubit by qubit, by means of partial state transfer circuits, which can be written as pairs of entangler and correlation eraser gates, similar to the $\ket{x}\ket{0}\rightarrow \ket{x}\ket{x}$ and $\ket{x}\ket{x}\rightarrow\ket{0}\ket{x}$ pair that appears in qubit state transfer. 

\begin{figure}[ht!]
\centering
\includegraphics{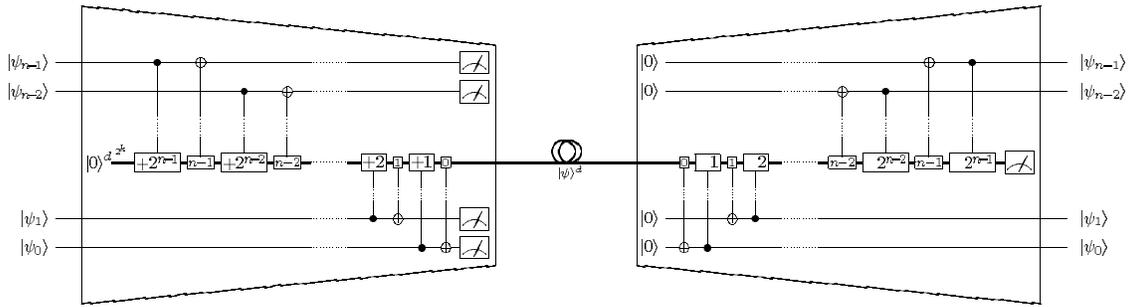}
\caption{Quantum multiplexer and demultiplexer schemes.\label{muxdemux}} 
\end{figure}

The gates of the circuit of Fig. \ref{binQuditST} can be put in terms of these partial transfer subblocks if the gates are reordered. This reordering is valid because some of the gates commute. The last ${C^{\mathcal{S}_{n-1}}X_{2^{n-1}}^{d}}^{\dag}$ will only have an effect on states $\ket{x}^d$ with $x\geq 2^{n-1}$. The input state for the lower line is $\ket{0}^d$. If we exclude the first gate, even all the other gates combined cannot increase the value of $\ket{0}^d$ up to that number. So, we can move the subtraction gate near to the $C^{\mathcal{S}_{n-1}}X_{2^{n-1}}^{d}$ gate. We can recursively apply this reasoning until we find a convenient reordering for the MUX gate. Looking back at the construction of Fig. \ref{muxgeneral}.c we can give the complete multiplexer and demultiplexer scheme for HDMA (Figure \ref{muxdemux}). 

Two types of gate are used: qubit controlled addition and subtraction modulo $d$ over the qudit, $CX^d_m$ and ${CX^d_m}^{\dag}$, and qudit controlled NOT, $C^{\mathcal{S}_ i}X^2_1$. The number $i$ on the qudit control line, for a qudit state $\ket{x}^d$, is the $i$-th digit of the binary representation of $x$. The right part of the circuit, on the receiver side, implements the inverse qudit transfer from $\ket{\psi}^d$ to the qudit formed by the global state of the $n$ qubits. 
 
The separation in partial state transfer gate pairs is important when it comes to the scaling of the solution. The same configurable block is valid for all the channels. An additional user can be added with an extra partial state transfer and an erasure at both sides of the transmission. If the qudit is a virtual collection of states from an infinite-dimensional space, as in optical angular momentum qudits, the qudit line can be kept and the system can accommodate a different number of users without a new qudit design.

What's more, each of the $CX^d_m$ gates acts only on the value of one of the digits of the binary expansion of $\ket{x}^d$, different for each partial transfer, and the effect on the qudit is ignored by all of the qudit controlled CNOT gates that are associated to the other qubits. This means that the multiplexing and demultiplexing blocks for different channels commute and we can add or recover a channel at any point of the transmission, instead of multiplexing or demultiplexing only at two fixed points. 

\section{Example.}
\label{eg}
\begin{figure}[ht!]
\centering
\includegraphics{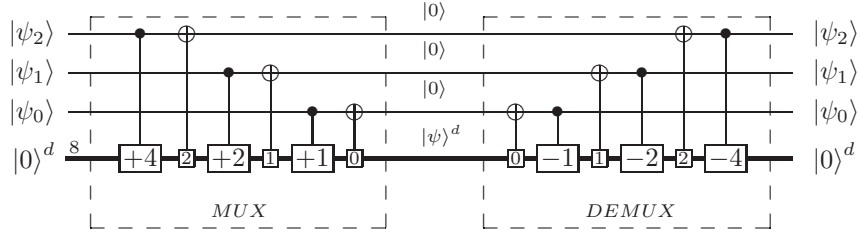}
\caption{Quantum multiplexer for 3 qubits.\label{mux3ex}} 
\end{figure}

Suppose a local network with three users that can only have one output to the network. Following the architecture of the previous section, the multiplexer circuit of figure \ref{mux3ex} transfers the state to the qudit line, and recovers the original channels at the receiver.  

The operation can be divided in partial state transfers for each of the channels. First we transfer the state of the first qubit onto the qudit (with $\ket{0}^8$ for $\ket{0}$, and $\ket{4}^8$ for $\ket{1}$).
\begin{equation}
\ket{\psi_2}\ket{\psi_1}\ket{\psi_0}\ket{0}^8 = (\alpha_2 \ket{0}+\beta_2 \ket{1})\ket{\psi_1}\ket{\psi_0}\ket{0}^8 \stackrel{CX^8_4}{\longrightarrow}  \alpha_2 \ket{0}  \ket{\psi_1}\ket{\psi_0}\ket{0}^8+\beta_2 \ket{1} \ket{\psi_1}\ket{\psi_0}\ket{4}^8. 
\end{equation}
The next gate will erase the correlation to the original qubit, $\ket{\psi_2}$, 
\begin{equation}
\stackrel{C^{\mathcal{S}_2}X^2_1}{\longrightarrow}  \alpha_2 \ket{0}  \ket{\psi_1}\ket{\psi_0}\ket{0}^8+\beta_2 \ket{0} \ket{\psi_1}\ket{\psi_0}\ket{4}^8=\ket{0}\ket{\psi_1}\ket{\psi_0} ( \alpha_2 \ket{0}^8+\beta_2 \ket{4}^8)=\ket{0} (\alpha_1 \ket{0}+\beta_1 \ket{1})\ket{\psi_0}(\alpha_2 \ket{0}^8+\beta_2 \ket{4}^8).
\end{equation}

The same process can be aplied to the qubit $\ket{\psi_1}$:
\begin{equation}
\stackrel{CX^8_2}{\longrightarrow} \ket{0}(\alpha_1\alpha_2\ket{0}\ket{\psi_0}\ket{0}^8 + \alpha_1\beta_2\ket{0}\ket{\psi_0}\ket{4}^8 +\beta_1\alpha_2\ket{1}\ket{\psi_0}\ket{2}^8 + \beta_1\beta_2\ket{1}\ket{\psi_0}\ket{6}^8).
\end{equation}
Now we have four states. For $\ket{2}$ and $\ket{6}$, 010 and 110 in binary, there is a 1 in the second position, while the others have a 0. So only in these cases, for which the qubit in $\ket{1}$, the state suffers a transformation (into $\ket{0}$) and the correlations are erased.
\begin{equation}
\stackrel{C^{\mathcal{S}_1}X^2_1}{\longrightarrow}  \ket{0}\ket{0}\ket{\psi_0}(\alpha_1\alpha_2\ket{0}^8 + \alpha_1\beta_2\ket{4}^8 +\beta_1\alpha_2\ket{2}^8 + \beta_1\beta_2\ket{6}^8).
\end{equation}

The partial state transfer from the $\ket{\psi_0}$ qubit completes the encoding of all the information into the qudit.
\begin{eqnarray}
\stackrel{CX^8_1}{\longrightarrow}  \ket{0}\ket{0}(&\alpha_0 \alpha_1\alpha_2\ket{0}\ket{0}^8& + \alpha_0 \alpha_1\beta_2\ket{0}\ket{4}^8 +\alpha_0 \beta_1\alpha_2\ket{0}\ket{2}^8 + \alpha_0 \beta_1\beta_2\ket{0}\ket{0}\ket{6}^8 \nonumber \\
+&\beta_0 \alpha_1\alpha_2\ket{1}\ket{1}^8& + \beta_0 \alpha_1\beta_2\ket{1}\ket{5}^8 +\beta_0 \beta_1\alpha_2\ket{1}\ket{3}^8 + \beta_0 \beta_1\beta_2\ket{0}\ket{1}\ket{7}^8).
\end{eqnarray}
Now, there are four qudit states that trigger the erasure, one for each state affected by the $\ket{1}$ part of the qubit.
\begin{equation}
\stackrel{C^{\mathcal{S}_1}X^2_1}{\longrightarrow}  \ket{0}\ket{0}\ket{0}(\alpha_2\alpha_1\alpha_0\ket{0}^8 +\alpha_2\alpha_1\beta_0\ket{1}^8 +\alpha_2\beta_1\alpha_0\ket{2}^8 +\alpha_2\beta_1\beta_0\ket{3}^8 +\beta_2\alpha_1\alpha_0\ket{4}^8 +\beta_2\alpha_1\beta_0\ket{5}^8 +\beta_2\beta_1\alpha_0\ket{6}^8 +\beta_2\beta_1\beta_0\ket{7}^8).
\end{equation}

After the measurement, we have a state that is equivalent to the original three qubits, $\ket{\psi}^{b}=\ket{\psi_2}\ket{\psi_1}\ket{\psi_0}=(\alpha_2 \ket{0}+\beta_2 \ket{1})\otimes (\alpha_1 \ket{0}+\beta_1 \ket{1})\otimes (\alpha_0 \ket{0}+\beta_0 \ket{1})$:

\begin{center}
\begin{tabular}{l@{=}l@{}r@{+}l@{}r@{+}l@{}r@{+}l@{}r@{+}l@{}r@{+}l@{}r@{+}l@{}r@{+}l@{}r}
$\ket{\psi}^b$&$\alpha_2\alpha_1\alpha_0$&$\ket{000}$&$\alpha_2\alpha_1\beta_0$&$\ket{001}$&$\alpha_2\beta_1\alpha_0$&$\ket{010}$&$\alpha_2\beta_1\beta_0$&$\ket{011}$&$\beta_2\alpha_1\alpha_0$&$\ket{100}$&$\beta_2\alpha_1\beta_0$&$\ket{101}$&$\beta_2\beta_1\alpha_0$&$\ket{110}$&$\beta_2\beta_1\beta_0$&$\ket{111},$\\
$\ket{\psi}^8$&$\alpha_2\alpha_1\alpha_0$&$\ket{0}^8$&$\alpha_2\alpha_1\beta_0$&$\ket{1}^8 $&$\alpha_2\beta_1\alpha_0$&$\ket{2}^8 $&$\alpha_2\beta_1\beta_0$&$\ket{3}^8 $&$\beta_2\alpha_1\alpha_0$&$\ket{4}^8 $&$\beta_2\alpha_1\beta_0$&$\ket{5}^8 $&$\beta_2\beta_1\alpha_0$&$\ket{6}^8 $&$\beta_2\beta_1\beta_0$&$\ket{7}^8.$
\end{tabular}
\end{center}

Once all of the input channels are put to $\ket{0}$, we can measure them to make sure the encoding happened without errors. If a $\ket{1}$ is found, there was an error.

Demultiplexing at the receiver is achieved by applying the inverse operations. $\ket{0}\ket{0}\ket{0}\ket{\psi}^8$, after the $C^{\mathcal{S}_0}X^2_1$ gate becomes, 
\begin{eqnarray}
\stackrel{C^{\mathcal{S}_0}X^2_1}{\longrightarrow}  \ket{0}\ket{0}(&\alpha_2\alpha_1\alpha_0\ket{0}\ket{0}^8& +\alpha_2\alpha_1\beta_0\ket{1}\ket{1}^8 +\alpha_2\beta_1\alpha_0\ket{0}\ket{2}^8 +\alpha_2\beta_1\beta_0\ket{1}\ket{3}^8 \nonumber \\
+&\beta_2\alpha_1\alpha_0\ket{0}\ket{4}^8& +\beta_2\alpha_1\beta_0\ket{1}\ket{5}^8 +\beta_2\beta_1\alpha_0\ket{0}\ket{6}^8 +\beta_2\beta_1\beta_0\ket{1}\ket{7}^8).
\end{eqnarray}
The ${CX^8_1}^\dag$ gate erases the traces of $\ket{\psi_0}$ in the qudit.
\begin{eqnarray}
\stackrel{{CX^8_1}^\dag}{\longrightarrow} \ket{0}\ket{0}(&\alpha_2\alpha_1\alpha_0\ket{0}\ket{0}^8& +\alpha_2\alpha_1\beta_0\ket{1}\ket{0}^8 +\alpha_2\beta_1\alpha_0\ket{0}\ket{2}^8 +\alpha_2\beta_1\beta_0\ket{1}\ket{2}^8 \nonumber \\
+&\beta_2\alpha_1\alpha_0\ket{0}\ket{4}^8& +\beta_2\alpha_1\beta_0\ket{1}\ket{4}^8 +\beta_2\beta_1\alpha_0\ket{0}\ket{6}^8 +\beta_2\beta_1\beta_0\ket{1}\ket{6}^8).
\end{eqnarray}

The four qudit states that remain can be extracted as a common factor and we have $\ket{0}\ket{0}(\alpha_0\ket{0}+\beta_0\ket{1})(\alpha_2\alpha_1\ket{0}^8+ \alpha_2\beta_1\ket{2}^8 + \beta_2\alpha_1\ket{4}^8 + \beta_2\beta_1\ket{0}\ket{6}^8)$. Notice that, at this point, channel 0 is again independent from the rest of the channels and from the qudit. After erasing the correlations, measuring the state of the qudit would destroy the information of the rest of the users but not channel 0. It is evident from the example that the destinations need not to be at the same point. We can think of a network where the qudit delivers each channel at a different point of its path to the last receiver. We can repeat the process until all of the original channels are recovered.

To get second qubit, we apply the second binary position controlled CNOT gate,
\begin{equation}
\stackrel{C^{\mathcal{S}_1}X^2_1}{\longrightarrow}  \ket{0}(\alpha_2\alpha_1\ket{0}\ket{\psi_0}\ket{0}^8 + \alpha_2\beta_1\ket{1}\ket{\psi_0}\ket{2}^8 +\beta_2\alpha_1\ket{0}\ket{\psi_0}\ket{4}^8 + \beta_2\beta_1\ket{1}\ket{\psi_0}\ket{6}^8).
\end{equation}
If we were to measure in this moment, the state of $\ket{\psi_1}$ would be affected by the qudit state, as they are still entangled. In order to avoid that, we erase the correlation.
\begin{equation}
\stackrel{{CX^8_2}^\dag}{\longrightarrow}  \ket{0}\ket{\psi_1}\ket{\psi_0}(\alpha_2\ket{0}^8  +\beta_2\ket{4}^8).
\end{equation}

We already have the last qubit encoded in the qudit state, with $\ket{0}^8$ corresponding to logical $\ket{0}$ and $\ket{4}^8$ corresponding to logical $\ket{1}$. The last two gates will transfer this state to a two-dimensional system, more adequate for the quantum gates at the destination quantum computer. Although it is not really necessary to measure the qudit state it can serve as an additional check for the correct operation of the system. If we find any state other than $\ket{0}^8$, there was some problem, like an eavesdropper or decoherence, during the transmission. 

\section{Discussion.} 
\label{discussion}
Starting from quantum state transfer, a quantum multiplexing scheme for multiple access quantum communications can be derived. HDMA broadens the options in quantum multiple access and gives a circuit for $n$ qubits to qudit and qudit to $n$ qubits conversion. Quantum multiplexers can be useful whenever such a conversion is needed. The two clearer examples are the communication of qubits between distant nodes in a quantum network and quantum buses inside a quantum computer. They are related communication problems, each one at a different distance scale.

In quantum networks we need a compact mean to take full advantage of the usually scarce communication resources. Remote quantum computers can be connected by intermediate qudit channels. The second problem is one of quantum computer architecture. In a complete quantum computer, quantum buses are needed to connect the different quantum registers, or to take the qubits from the quantum memory to the quantum processor. In both cases we can take advantage of the fact that during the pure communication step no information processing is needed.

Quantum information must be encoded into a physical system. From the desiderata of properties of such a system must have \cite{Div00}, there are two somewhat contradictory criteria. Quantum information units, be them qubits or qudits, should be resilient to decoherence and, at the same time, allow for the construction of efficient quantum gates. The loss of quantum superposition decoherence brings about can destroy one of the key features from which quantum computers get their exceptional computing properties. Without efficient, controlled interaction between qubits, no computation is possible. But, as a rule, the easier it is to have an interaction with a quantum system, the greater the effect of decoherence is.  

A good example is the one of photonic qubits. Photons are especially well suited for long distance quantum communication. The success of existing quantum optical networks, such as those used in quantum cryptography, owes much to the large coherence time of photons. Photons' resistance to decoherence stems from their particularly weak interaction with the environment and with other photons. This very same feature makes it difficult to build efficient photonic gates. This tradeoff between coherence time and gate construction is a constant in the search for physical systems to implement quantum computers

The existence of a quantum multiplexer permits to have one kind of information unit for the quantum operations, and another one for the transmission. Although some interaction is needed for the state transfer, the total number of gates is only four for each qubit, two at the multiplexer and two at the demultiplexer, and, then, the transfer of the information into a different system can be done with another encoding in a less sensitive to decoherence system. Later, a more interacting form of information can be fed into the quantum gates for the processing. 

\subsection{Hilbert space distribution of the channels' information.}
\begin{figure}[ht!]
\centering
\includegraphics{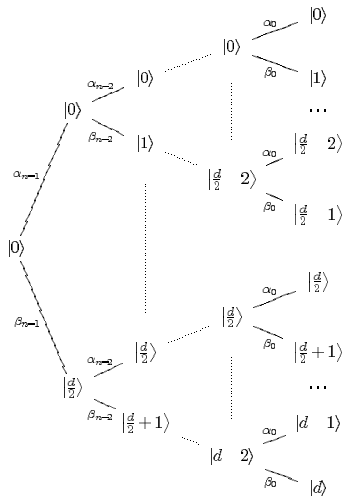}
\caption{Qudit state branching in the $2^n$-dimensional Hilbert space.\label{tree}} 
\end{figure}

The whole process can be seen as a redistribution of the probability amplitude in the qudit. In the first step $\ket{0}^d$ has a probability of 1. As the different qubits are encoded, the state is diffused through the Hilbert space, occupying $2^m$ states at the $m$-th step. Figure \ref{tree} shows the branching process that distributes the original information into the $d$-dimensional Hilbert space. Each column corresponds to a new partial transfer/erasure pair. The coefficients of each of the qudit states can be reconstructed by multiplying the qubit coefficients that are picked up on the branches on the way to the state.  

The final state will be
\begin{equation}
\ket{\psi}^d=\sum_{k=0}^{d-1} \left( \prod_{i=0}^{n-1} \alpha_i^{\bar{b_i}}\beta_i^{b_i} \right)\ket{k}^d,
\end{equation}
where $b_i$ is the $i$-th binary digit of $k$. The coefficients are the same of the binary expansion of $k$, with $\alpha$s in the place of 0s and $\beta$s in the place of 1s. 

This multiplexing can also be described in terms of subspaces of the whole qudit space. Each of the $n$ qubits is encoded in one half of the d-dimensional Hilbert space of the qudit. The state of the $i$-th qubit, $\ket{\psi_i}$, will be mapped into two orthogonal subspaces that divide the qudit space into two. There are two subspaces; states in $S_i^0$ represent $\ket{0}$ and states in $S_i^1$ indicate a qubit in $\ket{1}$, with  $S_i^0 \perp S_i^1$ and $ S_i^0 \cup S_i^1 = \mathcal{H}^d$, for $\mathcal{H}^d$ the global Hilbert space of dimension d. The subspaces are spanned by the states with a 0 or with a 1 in the $i$-th digit of the binary representation of the number of the state, so that $S_i^{0}=\{ \ket{x} :  x$ div $2^i$ is even$\}$ and $S_i^{1}=\{ \ket{x} :  x$ div $2^i$ is odd$\}$. 

At each multiplexing stage all the states are divided into two components, one for the new qubit's $\ket{0}$ and another one for the $\ket{1}$. The number of occupied states is doubled with each user. At the end of the multiplexing process, the information of the $n$ qubits is distributed among the $2^n$ states of the qudit. The demultiplexing combines the states again, transferring the probability amplitudes that gave the separation into the recovered qubit. Each state of the qudit target space can be seen as the intersection of the subspaces that have at their corresponding indices the binary representation of the state number, 
\begin{equation}
\ket{x}= \displaystyle \bigcap_{i=0}^{n-1} S_i^{b_i},
\end{equation}
for $b_{n-1}b_{n-2}\ldots b_0$ the binary string that represents $x$.

The probability amplitude associated with a channel's $\ket{0}$ and $\ket{1}$ states will be distributed through a different number of states depending on how many other channels there are. The mapping occurs in two steps. First, entanglement between the qubit and the target qudit subspace is created in the $2^{n+1}$-dimensional space of the qubit/qudit joint space. Later, the spaces are separated by erasing the correlations so that the reduced density matrix of the qudit system has the information formerly in the qubit. 

A similar spreading of one qubit state through a larger multidimensional subspace is found in multilevel encoding \cite{GBR06}. In this case, the spreading is a mean to fight decoherence. We can imagine certain forms of decoherence as a multiplexer whose inputs are a qubit channel and $n-1$ other channels representing the degrees of freedom of the environment that can couple with the qubit, in a line similar to that of the `caricature' model of decoherence as CNOT gates with the environment of \cite{Zur03}. If the subspaces are properly chosen, the unknown effect of the environment won't prevent us form recovering the qubit, in the same way that adding or dropping other channels at arbitrary points does not impede extracting the desired qubit from the qudit just by applying the corresponding partial state transfer and erasure pair of gates. Multilevel encoding can also give efficient gates to operate on the individual qubits while they are still in the qudit.

\subsection{Multiplexers in quantum communication.}
Up to this point, we have considered the $n$ qubit channels to be independent, and an intuitive product state interpretation arises. If the qubits are entangled, everything still holds, since, from the beginning, we have studied multiplexing as a qudit to qudit state transfer. The given circuits already take the $n$-qubit system as a whole. The proposed quantum multiplexer scheme is still valid in quantum netorks with applications where different channels can be entangled . From the basic quantum communication perspective, though, it will usually be easier to think of each channel as a separate entity, as entanglement does not usually appear on communication between different channels. When multipartite entanglement is needed, it is usually preferred to have an ealier stage of entanglement distribution, or flexible entanglement distribution networks \cite{HB04,BHM96}. These proposed networks usually have a central node that distributes correlations among the users. When multipartatite entanglement is needed entanglement swapping schemes are used \cite{ZHW97,BVK98}.

Many existing quantum information applications can be seen under a new light as a case of quantum multiplexing. In classical information transmission through a quantum network the erasure of the original qubit is not necessary. In that case, only the partial state transfer gates are needed at the transmitter. The final state will be only one of the $2^n$ possible values. Thus, at the receiver, we could skip the demultiplexing gates and proceed with a measurement, or include some $\mathcal{S}_i$ controlled NOT gates if we want to generate multiple qubits carrying the classical information. 

If we have two channels sending classical information, we only need a +2 and a +1 gate controlled by the first and second bit respectively. One possible embodiment of the qudit is a Bell state. Then, the contents will be sent as $\ket{0}^4=\frac{\ket{00}+\ket{11}}{\sqrt{2}}, \ket{1}^4=\frac{\ket{00}-\ket{11}}{\sqrt{2}},\ket{2}^4=\frac{\ket{10}+\ket{01}}{\sqrt{2}}$ or $\ket{3}^4 =\frac{\ket{10}-\ket{01}}{\sqrt{2}}$ for $\ket{0}\ket{0},\ket{0}\ket{1},\ket{1}\ket{0}$, and $\ket{1}\ket{1}$ respectively. The $CX^4_2$ gate corresponds to a CNOT on the second qubit of the Bell pair, and the $CX^4_1$ to a controlled sign shift, CZ, on the same qubit. The CZ operation is defined as $CZ\ket{x}=(-1)^x\ket{x}$, which changes the sign of $\ket{1}$ states. Both operations can be done without the presence of the first qubit. So, it is possible to send the first qubit in advance and include both channels in one qubit. This is exactly what happens in superdense coding. 

This economy of qubits cannot be extended for quantum information transmission. If a superposition exists, we must erase the correlations to the original qubit channels. Instead of the whole qudit, only one half of the system is available, and the state cannot be separated using the present qubits alone. Measuring the $n$ channels would destroy superposition. Keeping data in a quantum memory would separate the states in a larger Hilbert space, with effects akin to those of decoherence. As each qudit state would be entangled to different states of the ``environment'', here the quantum memory, interferences in the recovered qubits would be prevented. 

The two qubits must be interpreted as a four-dimensional system. For multiple systems, when no entanglement is present, each system can be studied separately, and the $n$ bits of classical information $n$ qubits carry can be asigned one bit to each qubit. If there is entanglement the $n$ bits of information are associated to the joint state \cite{BZ99}. In some cases, the information can even be distributed through the Hilbert space in such a way that no single system, on its own, can be said to carry one bit of information. In maximally entangled states, like the Bell pairs, the information is on the joint system. In superdense coding the information is encoded on the whole qudit. We can act on the encoding even if half of the system is not present, but at the price of loosing the capacity to send superpositions. 

\subsection{HDMA, orthogonality and multiple access.}

The motivation behind multiplexing is sharing a scarce communication resource, usually at the cost of an increase of complexity in another domain. To avoid interference between users, the signals that transmit the information must be orthogonal. 

In quantum information, qubits are given by systems with two orthogonal states that represent $\ket{0}$ and $\ket{1}$. The orthogonality, here, guarantees that there is no interference between the states. The schemes of classical multiple access can be used to achieve this orthogonality. Photonic qubits are quite illustrative to this respect. One example are time-bin qubits \cite{BGT99,TTT02,MRT03} and qudits \cite{SZG05}, where a pulse is divided into a superposition of two or $d$ pulses with a long enough temporal separation so that they can no longer interfere. This can be interpreted as a form of TDMA. A spectral separation similar to FDMA/WDMA has also be proposed for quantum information units \cite{Mol98,BGE99,LWE00} and there are experiments where qubits have been encoded in the sidebands of phase modulated light \cite{SMF95,MMG99,GMS03}. In dual rail and polarization encoded optical qubits \cite{Ral06}, the orthogonal modes are separated paths \cite{CY96,KLM00}, a form of SDMA, and different polarizations \cite{PJF02b,DRM03}, like in PDMA. In all those cases orthogonality is used as a separator for logical values. The same degrees of freedom that can be used for this separation can be employed to combine channels. 

There is also a plethora of photonic nonclassical states, each with its own special characteristics \cite{Lou00,Dod02}. Some examples are photon number states, or coherent and squeezed states. We also have at our disposal a great variety of operators that define the observables of photon modes, which present diverse uncertainty relations between non commuting observables. Quadrature operators, for instance, give analogues to position and momentum observables, and we can measure, among others, the number of photons, and the phase of the field. Each set of states has its own orthogonality properties that can serve as the base of a new multiple access scheme.

There exist many connections between qubit and qudit encodings and orthogonal states and classical multiplexing, modulation and symbol encoding. Different photon number states can encode qubits, like in single-rail encoding \cite{LR02,LLC03}, in a way not unlike the classical different voltage level assignements to represent different symbols. It is also possible to create qudits where the number of photons gives the state number, with photon number state $\ket{k}$ corresponding to the $\ket{k}^d$ qubit state, or with more complex correspondences. This representation could be the base of a HDMA system. In fact, a similar energy-based separation has been proposed for classical information as Power Division Multiple Access \cite{Maz98}. 

Continuous signals can be decomposed into in-phase and quadrature components. This decomposition is at the origin of classical quadrature modulation schemes, such as the modulation with multilevel symbols of QAM (quadrature amplitude modulation), and gives related quadrature multiplexing strategies \cite{Hay01}. Quantum information with continuous variables is usually based on the quadratures of a mode. With these canonical operators the methods of classical communications can be employed \cite{BV05,FSB98,SKO06}. Qubits and qudits can be encoded into the infinite-dimensional Hilbert space of the canonical coordinates of photon modes \cite{PMV04}, opening the door for usage in multiple access. Other classical modulation-related strategies, like multicarrier modulation multiple access based on OFDM (orthogonal frequency division multiplexing) \cite{Bin90, ADL98}, are much used in wideband systems and modern wireless networks. A quantum extension could be derived relating different subspaces to the various subcarriers.  

There are also schemes that encode qubits in two near orthogonal coherent states \cite{CMM99,RGM03,GVR04}. As there are infinite coherent states with an almost null overlap, they seem fit to quantum multiplexing scenarios where the same channel can be used for any number of users, choosing a new coherent state when a new user joins the channel. This multiple access philosophy can mimic classical CDMA networks with channel overloading. In those networks, almost orthogonal codes permit to accommodate an increasing number of users, at the cost of a degradation in the transmission quality that is proportional to the user number at a particular time \cite{YKK00,SVM00}. The greater the number of users, the more difficult it becomes to find states with a small overlap. It would also be interesting to reproduce the same behaviour with an imperfect state transfer circuit that maps $n$ qubits into a $d< 2^n$ qudit. The given quantum multiplexer has some common features with certain forms of channel coding and quantum data compression \cite{CM00,Lan02,BHL06}. Methods for efficient channel coding could be adapted for multiplexing to give multiple access scenarios with a flexible upper bound on the number of users.

Optical angular momentum, OAM, is also a promising incarnation for qudits \cite{TDT03,VPJ03,MVR04} and a good candidate for optical multiplexing, carrying the information of many channels in a single photon. Optical angular momentum in optical vortices has already been used for multilevel classical communication \cite{BC04,GCP04,RHB06}.

More generally, any observable with many incompatible outcomes can give rise to new multiple access schemes. The principle is the same as the one of qubit design, finding orthogonal states. Other systems will have their own observables, equivalent or not to photons' operators. 

\subsection{Overview on HDMA}

HDMA can be seen as a generalization that encompasses all other multiple access schemes. HDMA has a strong connection with CDMA. CDMA is the prevailing multiple access model in third generation mobile radio networks, and has established itself as one of the most important classsical multiple access technologies. In CDMA, users exploit the orthogonality, or near orthogonality, of certain codes to share the same band of the spectrum at the same time \cite{PSM82,PO98,Sta01}. Among other advantages, CDMA offers an improved behaviour against noise and a dynamical limit for the number of users, with a graceful degradation when the number of channels that can be perfectly separated has been exceeded. TDMA and FDMA can be put as special cases of CDMA. The formalism of CDMA makes use of the orthogonality of signals in the Hilbert space. The main different between the CDMA and HDMA is the point of view. HDMA stresses some aspects that are often overlooked in the standard formulation of CDMA, like the complex nature of the elements of the Hilbert space, and the interpretation of coding and decoding in terms of subspace projections. In our multiplexer there are also considerations that are not taken in classical systems. The first point of divergence is the presence of entanglement, which makes it necessary to erase correlations to be able to dispose of the information in the qubits. The second difference is the no-cloning prohibition in quantum information, which difficults the recovery at the receiver. Instead of having one copy for each output channel, we need a sequence of gates to extract the information for each destination and erase the correlations with the common channel (the qudit). 

The treatment under the HDMA framework unifies the study of multiple access techniques and can be a useful abstraction when comparing systems. It is common to find qudits that combine orthogonality in more than one domain. Some examples are transverse momentum and position entangled qudits \cite{OKB05}, single photon implementations of two qubits using spatial and polarization modes \cite{EKW01} or hyperentangled photons, with polarization, spatial mode, and time energy orthogonality \cite{BLP05}. This also happens in classical systems, where hybrid multiple access methods appear and different domains are used in the separation of uplink and uplink channels, or \emph{duplexing}, and the separation of channels in multiplexing \cite{Rap01}. HDMA emphasizes the equivalence of these systems with others with qubits implemented only in the temporal (TDMA) \cite{TAZ04}, spatial (SDMA) \cite{NPS04,NLA05} or optical angular momentum \cite{MVW01} degree of freedom.

Multiplexing is a process of  re-encoding, and can be understood as a state transfer. The same information is put together in a channel that exhausts the more scarce resource, at the cost of another resource that is not so limited. This tradeoff appears in all multiple access systems. For instance, in optical networks, where fibre links are a precious resource, multiplexers take many channels into one link avoiding the costly deployment of optical fibre. There are multiple equivalent systems for qudits. The more trivial embodiement of a $d=2^n$ qudit is $n$ qubits in $n$ channels. For information transmission we prefer more economical encodings, like OAM qudits that can send $n$ qubits in a single photon without increasing the time span or the number of spatial modes. A new important resource in quantum communications is the coherence time. Inside each quantum computer, relatively easy interacting qubits are desirable to perform the quantum algorithms, but, during the transmission, decoherence resilient qudits are better, as the only interactions occur at the multiplexer and demultiplexer, and maybe inside quantum repeaters. Other issues that should be taken into account include error correction, and fault tolerance. A single qudit is more vulnerable to photon loss problems, but with error correction codes \cite{Got96,CS96} this and other problems can be overcome. 

We have presented a quantum version the multiplexer and demultiplexer gates that appear in communication networks and in the inner architecture of quantum computers. These circuits can be better understood in terms of a qudit state transfer, where the particularities of quantum information are taken into account. The explicit circuit is formed by a group of $n$ subunits that define the inclusion and extraction of each user into the channel and erase the contents of the qubit to avoid undesired entanglement. The HDMA scheme gives a framework to compare different multiple access proposals and to derive new multiple access methods. With it, it is possible to gain some insight into qubit state separation and coding issues. Hopefully, HDMA will improve the potential of new quantum networks where the quantum information of many users will be transmitted sharing the existing network resources for a more efficient distribution. 


\newcommand{\noopsort}[1]{} \newcommand{\printfirst}[2]{#1}
  \newcommand{\singleletter}[1]{#1} \newcommand{\switchargs}[2]{#2#1}

\end{document}